\documentclass[a4paper,12pt]{article}
\usepackage{graphicx}
\usepackage{amssymb}
\usepackage{amsmath}
\usepackage{cite}
\usepackage{axodraw4j}
\def\be{\begin{equation}}
\def\ee{\end{equation}}
\def\bea{\begin{eqnarray}}
\def\eea{\end{eqnarray}}
\def\beb{\begin{eqnarray*}}
\def\eeb{\end{eqnarray*}}

\newlength{\myVSpace}
\begin{document}
\begin{center}
{\Large {\bf Muon anomalous magnetic moment in the standard model extension}}
\vskip 4em{ {\bf$S.Aghababaei^{*}$} \footnote{s.aghababaei@ph.iut.ac.ir}\: ,
	\: {\bf$M.Haghighat^{\dag}$}\footnote{m.haghighat@shirazu.ac.ir}\: ,
	\: {\bf$I.Motie^{\ddagger}$}}
\vskip 1cm

$^*$ Department of Physics, Isfahan University of Technology, Isfahan
84156-83111, Iran\\
$^\dag$ Department of Physics, Shiraz University, Shiraz 71946-84795, Iran\\
$\ddagger$ Department of Physics, Mashhad Branch, Islamic Azad University, Mashhad, Iran
\vskip 1.9cm
{\bf Abstract}
\\

\medskip
\begin{minipage}{135mm}
{We consider the standard model extension to explore the anomalous magnetic dipole moment of the muon.  In the QED part of the theory for the CP and CPT-even Lorentz parameter $c_{\mu\nu}$, all independent electromagnetic form factors depend on a new scalar as $p'.c.p$.   Therefore, the form factors, even in zero momentum transfer, can be energy dependent.  We examine the magnetic form factor to find such an energy dependent up to the one loop level at the leading order of  $c_{\mu\nu}$.  We show that at the high energy limit (but low enough to satisfy $p^2/m^2\ll1$) there is an enhancement on the muon anomalous magnetic moment.  For the first time, we find a bound on the $c_{\mu\nu}$ components for the muon as $[c_{TT}+0.35(c_{XX}+c_{YY})+0.28c_{ZZ}]$, which is about $10^{-11}$ in a terrestrial experiment.}
\end{minipage}
\end{center}
\medskip
\section{Introduction}
The recent measurement on the muon anomalous magnetic dipole moment ($\mu$-AMDM) in the E821 experiment at Brookhaven National Laboratory \cite{E-821,BNL} has been provided a new place to study the standard model (SM) of particle physics and new physics beyond the SM.  In fact, the muon g-2 Collaboration has found a discrepancy above the $3\sigma$ level for the muon anomalous magnetic moment ($a_{\mu}$) with the SM prediction as \cite{a-exp}
\begin{eqnarray}
\Delta a_{\mu}=(a_{\mu})_{EXP}-(a_{\mu})_{SM}=(26.1\pm8.0)\times10^{-10}.\label{aexp}
\end{eqnarray}
In order to understand the difference between the SM prediction and the experimental measurement, many works in both theoretical \cite{theg-2} and experimental \cite{expg-2} aspects of $\mu$-AMDM have been done.  If we believe that the theoretical calculation within the standard model is complete, then this deviation should reflect the incompleteness of the SM and the presence of new physics beyond the standard model \cite{NP}. However, the new physics can be introduced by new interactions and/or new particles.  For instance, there are many attempts to calculate $a_{\mu}$ in the noncommutative space-time geometry \cite{NCg-2}, extra dimensional models \cite{exdg-2}, little Higgs model \cite{lhiggsg-2}, minimal supersymmetric standard model \cite{susyg-2}, and dark photon \cite{darkg-2} in which the deviation in $a_{\mu}$ is explained by introducing a new particle through an extra U(1) gauge boson beyond the ordinary photon. In this study, we would like to consider the so-called standard model extension (SME) which is a minimal extension of the standard model with  Lorentz symmetry violation terms.\\
Although, at low energy the Lorentz and CPT symmetries seem to be the exact symmetries of nature, the local Lorentz invariance at the Planck scale can be broken through quantum gravity. In fact, irrespective of the underlying
fundamental theory, the SME Lagrangian as an effective field theory has been introduced to containing such symmetry violation in the standard model\cite{SME}.  The presence of Lorentz violating (LV) terms in the SME can be induced by some appropriate  Lorentz  spontaneous symmetry breaking in a fundamental theory\cite{SSB}.  Therefore, these terms respect the observer Lorentz symmetry  while the particle Lorentz symmetry is violated.  Furthermore, the Lorentz and CPT symmetries have some relations through the CPT theorem in a local field theory \cite{CPT}, which also makes SME a suitable framework also for investigating the violation of the CPT symmetry. However, many works have been done on the theoretical and the phenomenological aspects of the SME \cite{LV} where terrestrial \cite{terrestrialSME} and astrophysical \cite{astrophysicalSME} systems have lead to restricted bounds on the LV parameters  \cite{table}.  In this article, we consider the QED part of SME (QEDE) to examine the appropriate LV parameters which can affect the AMDM of particles and obtain the corresponding modified form factors in the presence of the LV backgrounds. \\ 
In Sec. II, we introduce the QED part of SME.  In Sec. III, we examine $\mu$-AMDM in the QEDE.  For this purpose, the electromagnetic current for a charged fermion can be written in terms of appropriate form factors.  Then in this section, we obtain the magnetic form factor up to one loop level at the leading order of $c_{\mu\nu}$.  Consequently, $\mu$-AMDM is derived in the sun-centered inertial frame to find some bounds on the corresponding components of $c_{\mu\nu}$.  We give some concluding remarks in Sec. IV.  

\section{QED part of SME}
In the QED part of the SME charged fermions interact with photons in the four dimensions as follows \cite{SME}:
\begin{eqnarray}
\mathcal{L}^{QEDE}=\bar{\psi}(i\Gamma_{\mu}\overleftrightarrow{D}^{\mu}-M)\psi,\label{lqed}
\end{eqnarray}
where $D_{\mu}$ is the usual covariant derivative in QED, $\psi$ is a fermion field with mass $m$, and  
\begin{eqnarray}
\Gamma_{\mu}&=&\gamma_{\mu}+c_{\mu\nu}\gamma^{\nu}-d_{\mu\nu}\gamma^{\nu}\gamma^{5}
+e_{\mu}+if_{\mu}\gamma^{5}+\frac{1}{2}g_{\lambda\nu\mu}\sigma^{\lambda\nu},\nonumber\\
M&=&m+a_{\mu}\gamma^{\mu}-b_{\mu}\gamma^{\mu}\gamma^{5}
+\frac{1}{2}H_{\mu\nu}\sigma^{\mu\nu}+im_{5}\gamma^{5},\label{gamma0}
\end{eqnarray}
where the new parameters in $\Gamma_{\mu}$ and $M$ are called LV parameters.  The momentum-like parameters which appeared in $\Gamma_{\mu}$ are important at high energies while the mass-like parameters in $M$ are more effective at the low energies.  In fact, at the high energy limit one can safely ignore all the LV parameters in $M$.  Therefore, for a CP invariance quantity such as the magnetic dipole moment (MDM) we can consider the only the CP-invariance LV parameter in $\Gamma_{\mu}$ (i.e., $c_{\mu\nu}$) to rewrite the effective Lagrangian as follows:
\begin{eqnarray}
\mathcal{L}^{MDM}&=&\frac{i}{2}\bar{\psi}\gamma^{\mu}\overleftrightarrow{D}_{\mu}\psi-m\bar{\psi}\psi+\frac{i}{2}c_{\mu\nu}\bar{\psi}\gamma^{\mu}\overleftrightarrow{D}^{\nu}\psi,\label{lmdm}
\end{eqnarray}
which leads to a free field Dirac equation as
\begin{eqnarray}
(\not \! p-m+c_{\mu\nu} p^{\nu}\gamma^{\mu})u(p)=0,\label{dirac}
\end{eqnarray}
and a new Feynman rule for the fermion-photon vertex as
\begin{eqnarray}
\Gamma^{\mu}=-ie(\gamma^{\mu}+c^{\nu\mu}\gamma_{\nu}),\label{vertex}
\end{eqnarray}
where the Dirac gamma algebra at the same order is
\begin{eqnarray}
\{\Gamma_{\mu},\Gamma_{\nu}\}=\{\gamma_{\mu},\gamma_{\nu}\}+4c_{\mu\nu}^{S}.
\end{eqnarray}	
In fact, the Dirac algebra depends only on the symmetric part of $c_{\mu\nu}$ which leads to the independence of physical quantities to the antisymmetric part of $c_{\mu\nu}$.  Meanwhile, for a fermion propagator up to the first order of $c_{\mu\nu}$ one has 
\begin{eqnarray}
S_{F}(p)&=&\frac{i}{\not \!p-m+c_{\mu\nu} p^{\nu}\gamma^{\mu}},\nonumber\\
&=&\frac{i(\not \!p+m)}{p^{2}-m^{2}}-\frac{2i c^{S}_{\mu\nu}p^{\nu}p^{\mu}(\not \!p+m)}{(p^{2}-m^{2})^2}+\frac{ic_{\mu\nu}p^{\nu}\gamma^{\mu}}{p^{2}-m^{2}},\label{propagator}
\end{eqnarray}
and the Gordon identities at the leading order would modify to 
\begin{eqnarray}
\bar{u}(p')\gamma_{\mu}u(p)=\bar{u}(p')\bigg (\frac{(p+p')_{\mu}}{2m}+i\frac{\sigma_{\mu\nu}q^{\nu}}{2m}+i\frac{c^{\alpha\nu}\sigma_{\mu\alpha}q_{\nu}}{2m}+\frac{c_{\mu\nu}(p+p')^{\nu}}{2m}\bigg)u(p),\label{gordon1}
\end{eqnarray}
and
\begin{eqnarray}
\bar{u}(p')\gamma^{\nu}c_{\nu\mu}u(p)=\bar{u}(p')\bigg(\frac{(p+p')^{\nu}}{2m}+i\frac{c_{\nu\mu}\sigma^{\nu\alpha}q_{\alpha}}{2m}\bigg)u(p),\label{gordon2}
\end{eqnarray}
where $p$ and $ p'$ are momenta of the ingoing and outgoing fermions, respectively. For example, in the nonrelativistic limit the coupling of the current given in (\ref{gordon1}) with an electromagnetic vector potential leads to a Hamiltonian as 
 \begin{eqnarray}
 (i\frac{\sigma_{\mu\nu}q^{\nu}}{2m}+i\frac{c^{\alpha\nu}\sigma_{\mu\alpha}q_{\nu}}{2m}) \times (-eA^{\mu}),
 \label{A1}
 \end{eqnarray}
 where, for a magnetic field in the $z$ direction, the gauge independent part of the Hamiltonian can be cast into  
  \begin{eqnarray}
  -ie\frac{\sigma_{3}q_{2}A_{1}}{2m}+ie\frac{\sigma_{3}q_{1}A_{2}}{2m}
  -ie\frac{c_{22}\sigma_{3}q_{2}A_{1}}{2m} +ie\frac{c_{11}\sigma_{3}q_{1}A_{2}}{2m},
  \label{A2}
  \end{eqnarray}
or after a little algebra
   \begin{eqnarray}
  e\frac{1}{4m}\bigg (2+2\dfrac{(c_{11}-c_{22})}{2}\bigg )\sigma_{3}B_{3}.\label{treelevelgfactor}
   \end{eqnarray}
The first term in (\ref{treelevelgfactor}) shows the Dirac value for the g-factor while the rest terms are corrections in the QEDE at the tree level.\\  
  In the next section, we will explore the Lorentz violation effects on the g-factor up to the one loop.  Meanwhile, the other useful equations and identities at the first order of $c_{\mu\nu}$ can be found in Appendix A.

\section{$\mu$-AMDM in the QEDE}
In QED  for each $\frac{1}{2}$-spin point particle, the g-factor is 2  or $a=0$ at the tree level which is different from the experimental value given in (\ref{aexp}) for muon.  In fact, for the electron the QED loop corrections can completely explain the current experimental value for the anomalous magnetic moment of the electron $a_e$.  Meanwhile, the QED alone cannot explain the experimental value of  $a_{\mu}$ and one not only needs to consider the loop corrections through the SM framework but also the contribution from new physics beyond the standard model as
\begin{eqnarray}
a_{\mu}=a_{\mu}^{SM} + a_{\mu}^{NP},
\end{eqnarray}
where $a_{\mu}^{NP}$ contains all effects on the anomalous magnetic moment of muon from physics beyond SM and  
\begin{eqnarray}
a_{\mu}^{SM}=a_{\mu}^{QED}+a_{\mu}^{Weak}+a_{\mu}^{hadron},
\end{eqnarray}
where the $a^{QED}_{\mu}$  includes the Schwinger result \cite{Schwinger,Foley} plus corrections up to five loops, $a^{Weak}_{\mu}$ shows the weak contribution with the loops containing the heavy bosons $W^{\pm}$, Z, and H and the hadronic part  $a^{hadron}_{\mu}$ shows the contribution of hadrons in the loop corrections\cite{SM}.  For instance, in  (\ref{treelevelgfactor}) the effect of QEDE as a theory beyond SM on $a_{\mu}$ can be derived as
      \begin{eqnarray}
    a_{\mu}^{tree level}=\dfrac{g-2}{2}=\dfrac{(c_{11}-c_{22})}{2},
    \end{eqnarray}
 where this correction at the tree level of QEDE depends on the LV parameter $c_{\mu\nu}$ which leads to a bound on 
   $ c_{11}-c_{22}$ as order of $ 10^{-10} $ when compared with $a_{\mu}^{exp}$.  In the next subsections, we are going to explore the one loop correction on $a_{\mu}$ within the QEDE framework to find  $a_{\mu}^{NP}$ in this theory.
\subsection{Electromagnetic form factors in QEDE}
In QED, to study loop effects in the electromagnetic interaction of fermionic point particles usually currents are parametrized in terms of electromagnetic form factors.  In fact,  the vector current is a Lorentz vector and can be generally expanded in terms of all independent Lorentz vectors in the system under consideration.  However, $c_{\mu\nu}$ in QEDE is a new Lorentz quantity under observer Lorentz transformation which should be considered along with the other Lorentz vectors such as $\gamma_{\mu}$ and the momenta of fermions.  Therefore, the most general form of the fermionic  current, which is allowed by the Lorentz invariance and Ward identity, can be written as 
\begin{eqnarray}
\left\langle J_{\mu}^{em}\right\rangle =\overline{u}(p') {\cal{F_\mu}}(q^2)u(p),\label{current}
\end{eqnarray}
where $q=p-p'$ is the momentum transfer, 
\begin{eqnarray}
{\cal{F_\mu}}(q^2)&=&F_1\Big[\,\gamma_\mu+\gamma^\nu c_{\nu\mu}\Big]+F_2\Big[\,\,i\frac{\sigma_{\mu\nu}q^\nu}{2m}\Big]+(F_c)_{\mu},\label{formfactor}
\end{eqnarray}
and $F_1$ and $F_2$ are the electric charge and magnetic moment form factors, respectively, and depend on the Lorentz scalars such as $q^2$, $p_{\mu}c^{\mu\nu}p'_{\nu}$ and so on.  Meanwhile, the new Lorentz quantity $c_{\mu\nu}$ leads to new form factors which are collected in $(F_c)_{\mu}$ and can be defined as
\begin{eqnarray}
(F_c)_{\mu}&=&F_{c_1}\Big[\,\,(q.c.\gamma)q_{\mu}-q^{2}c_{\mu\alpha}\gamma^{\alpha}\Big]+iF_{c_2}\Big[\,c_{\mu\alpha}\sigma^{\alpha\nu}-c^{\nu\alpha}\sigma_{\alpha\mu}\Big]\frac{q_{\nu}}{2m},\label{formfactor1}
\end{eqnarray}
in which the new form factors $F_{c_1}$ and $F_{c_2}$ depend only on the scalar quantity $q^2$ at the leading order. In fact, up to the first order of $c_{\mu\nu}$ only the electric  and magnetic form factors $F_1$ and $F_2$ can have some dependence on the $c_{\mu\nu}$ through the scalar quantity $p_{\mu}c^{\mu\nu}p'_{\nu}$.  Therefore, values of the magnetic form factors can be enhanced at the higher energies even in a zero momentum transfer.  Before proceeding, some comments are in order.  The form factors $F_1$ and $F_2$ can also depend on the other LV parameters if one considers the whole Lagrangian $\mathcal{L}^{QEDE}$ given in (\ref{lqed}) and (\ref{gamma0}). However, such dependencies on the LV parameters can be categorized as follows:\\
(1) The LV parameters without any Lorentz indices.  For instance, $m_5$ in (\ref{gamma0}) is a Lorentz scalar with a mass dimension.  In this case, the dimensionless form factors can depend on $m_5/m_f$ where $m_f$ is the fermion mass that leads to a very small correction without any enhancement.\\
(2) The LV parameters with one Lorentz index such as $a_\mu$, $b_\mu$, and so on.  The Lorentz scalars that are formed with such parameters and momenta at the lowest order contain only one momentum vector.  However, the space part of such scalars usually averages out to zero.  For instant, in the storage ring of the E821 experiment, the rotating particle in the XY plane has zero average momentum and therefore the scalar such as $a_\mu p^\mu$ reduces to $a_0p_0$.  In this case, the dimensionless quantity is $a_0p_0/m_f^2$, which is also very small but enhances as $p_0/m_f$ for $p_0> m_f$.\\
(3) The LV parameters with two Lorentz indices.  In this case, there are three parameters where $c_{\mu\nu}$ and $d_{\mu\nu}$  are dimensionless and $H_{\mu\nu}$ has the dimension of mass. Therefore, the dimensionless scalars are $c_{\mu\nu}p^\mu p^\nu/m_f^2$, $d_{\mu\nu}p^\mu p^\nu/m_f^2$, and $\frac{H_{\mu\nu}}{m_f}p^\mu p^\nu/m_f^2$ where, for the antisymmetric tensor $H_{\mu\nu}$ and for an experiment like E821, the value of $\frac{H_{\mu\nu}}{m_f}p^\mu p^\nu/m_f^2$ averages out to zero for $\mu\ne\nu$.  Meanwhile, the CP-conserving form factors $F_1$ and $F_2$ cannot depend on the LV parameter $d_{\mu\nu}$ at the lowest order.  In fact, in a loop correction in the extended QED framework, to have a CP-invariant combination of $d_{\mu\nu}$, $\gamma_5$ and the momenta four-vectors, at least two vertices with the LV parameter $d$ are needed.  Therefore, the $d$ dependence of the form factors are proportional to  $(d_{\mu\nu}p^\mu p^\nu/m_f^2)^2\ll c_{\mu\nu}p^\mu p^\nu/m_f^2$ for $c\sim d$.\\
(4) Finally, the LV parameter $g_{\lambda\nu\mu} $ with three Lorentz indices leads to a Lorentz scalar as $g_{\lambda\nu\mu}p^\lambda p^\nu p^\mu/m_f^3 $, which results in a null value if one takes into account the zero average of momenta and the traceless  property of $g_{\lambda\nu\mu} $.

It should be noted that none of the LV parameters that are introduced in (\ref{gamma0}) have any contribution on the leading order corrections to the muon g-factor.  In fact, these parameters, as is shown in Ref. \cite{lowenergy}, affect the spin precession frequency through a combination of the parameters $H_{\mu\nu}$ and $g_{\lambda\nu\mu} $ that couple directly to $\sigma^{\mu\nu}$ and the parameters $b_\mu$ and $d_{\mu\nu}$.  Meanwhile, 
the LV parameters can directly correct the muon g-factor via their effects on the form factors $F_1$ and $F_2$.  However, the largest effect at the high energy limit comes from the $c_{\mu\nu}$ parameter, which is negligible in the  experiments for measuring the anomalous magnetic moment of fermions where $p^2/m^2\ll 1$.  Therefore, in the following sections, we examine the contribution of  $c_{\mu\nu}$ on the g-factor of muon at the high energy limit ($p^2/m^2\gg 1$).    

\begin{figure}
\centering
\includegraphics[scale=.6]{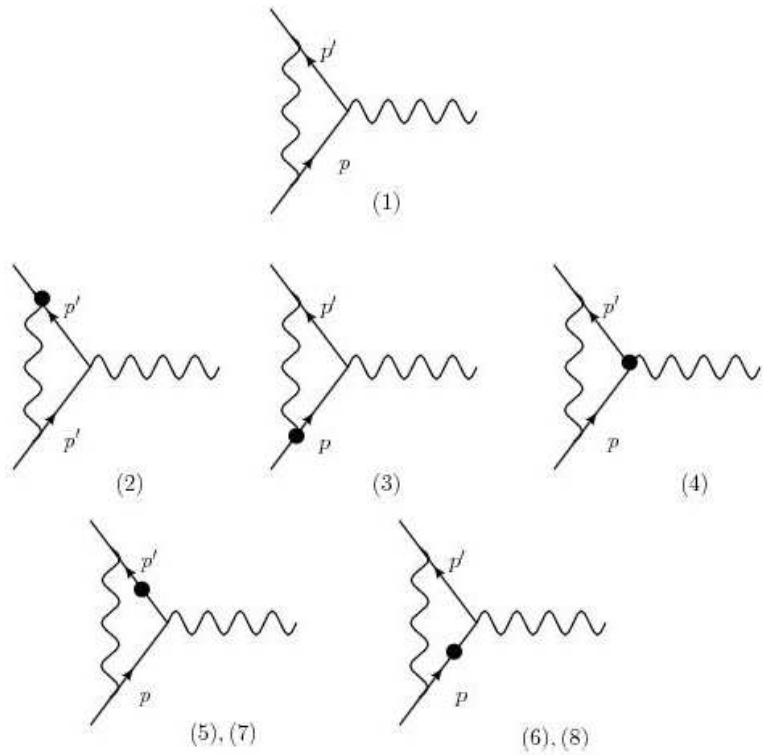}
\label{fig:digraph}
\caption{Vertex correction in the LV background. The bold circle shows where the LV background affect the vertex. In (1) only the wave functions have been changed.  In (2)-(4) the vertices have been corrected while the last two diagrams are devoted to the corrections on the fermion propagators.}\hskip -50cm
\end{figure}
\subsection{The muon vertex function in the LV theory}
The form factors given in (\ref{formfactor}) can be calculated perturbatively in the context of QEDE.  At the lowest order of the LV parameter only $F_1$ and $F_2$ depend on the  $c_{\mu\nu}$ through $p_{\mu}c^{\mu\nu}p'_{\nu}$ as a Lorentz scalar.  To explore such a dependence, we consider the fermion-photon vertex in the QEDE framework up to the one loop correction and the first order of the $c$ parameter.  To this end, we consider (\ref{dirac}), (\ref{vertex}), and (\ref{propagator}) to draw  six diagrams at the first order of $c$, as is shown in Fig.1.  
In fact, in the leading order, one can use (\ref{dirac}) to (\ref{propagator}) to evaluate the one loop correction in the QEDE as
\begin{eqnarray}
\delta\Gamma^{\mu}&=&\int\frac{d^{4}l}{(2\pi)^4}\int\frac{dxdydz\delta(x+y+z-1)}{(l^{2}-\Delta)^{3}}(-2ie^{2})\bar{u}(p')\{\mathcal{N}_1+...+\mathcal{N}_6\}u(p)\nonumber\\
&-&\int\frac{d^{4}l}{(2\pi)^4}\int\frac{dxdydz\delta(x+y+z-1)}{(l^{2}-\Delta)^{4}}(12ie^{2})\bar{u}(p')\{\mathcal{N}_7+\mathcal{N}_8\}u(p),\label{vf}
\end{eqnarray}
where $x$, $y$, and $z$ are the Feynman parameters, $\Delta=-xyq^{2}+(1-z)^{2}m^{2}$, $u(p)$ is the ordinary free Dirac spinor  except in the $\mathcal{N}_1$-term, and 
\begin{eqnarray}
\mathcal{N}_1&=&\gamma_{\rho}(\not \!k'+m)\gamma^{\mu}(\not \!k+m)\gamma^{\rho},
\end{eqnarray}
\begin{eqnarray}
\mathcal{N}_2&=&\gamma_{\rho}(\not \!k'+m)\gamma^{\mu}(\not \!k+m)c^{\alpha\rho}\gamma_{\alpha},
\end{eqnarray}
\begin{eqnarray}
\mathcal{N}_3&=&\gamma_{\rho}(\not \!k'+m)c^{\alpha\mu}\gamma_{\alpha}(\not \!k+m)\gamma^{\rho},
\end{eqnarray}
\begin{eqnarray}
\mathcal{N}_4&=&c_{\alpha\rho}\gamma^{\alpha}(\not \!k'+m)\gamma^{\mu}(\not \!k+m)\gamma^{\rho},
\end{eqnarray}
\begin{eqnarray}
\mathcal{N}_5&=&\gamma_{\rho}(\not \!k'+m)\gamma^{\mu}c_{\alpha\beta}k^{\beta}\gamma^{\alpha}\gamma^{\rho},
\end{eqnarray}
\begin{eqnarray}
\mathcal{N}_6&=&\gamma_{\rho}c_{\alpha\beta}k'^{\beta}\gamma^{\alpha}\gamma^{\mu}(\not \!k+m)\gamma^{\rho},
\end{eqnarray}
\begin{eqnarray}
\mathcal{N}_7&=&x\gamma_{\rho}(\not \!k'+m)\gamma^{\mu}c_{\alpha\beta}k^{\alpha}k^{\beta}(\not \!k+m)\gamma^{\rho},
\end{eqnarray}
\begin{eqnarray}
\mathcal{N}_8&=&y\gamma_{\rho}c_{\alpha\beta}k'^{\alpha}k'^{\beta}(\not \!k'+m)\gamma^{\mu}(\not \!k+m)\gamma^{\rho}.
\end{eqnarray}
 It should be noted that (\ref{vf}), besides the LV corrections at the leading order, contains all corrections up to the one loop level for the vertex function coming from the SM as well.  Therefore, one can consider (\ref{vf}) as 
\begin{eqnarray}
\delta\Gamma^{\mu}&=&\delta\Gamma^{\mu}_{SM}+\delta\Gamma^{\mu}_{LV},
\end{eqnarray}
where the first term is the usual SM correction and the last one shows the LV contribution on the vertex function as are given, respectively, in (\ref{SM}) and (\ref{LV}) in Appendix B. Now one can compare (\ref{vf}) with (\ref{formfactor}) to find all form factors $F_1$ to $F_c$ as are derived in Appendix C in (\ref{F1}), (\ref{F2}) and (\ref{Fc}), respectively.  For instance, the magnetic form factor besides the ordinary SM part has some contribution from the LV part of QEDE as 
\begin{eqnarray}
F^{LV}_2(q^2)&=&\int\frac{d^{4}l}{(2\pi)^4}\int\frac{dxdydz\delta(x+y+z-1)}{(l^{2}-\Delta)^{4}}\nonumber\\
&&(12ie^2)\bigg[[(1-z)z^2)(p.c^S.p)][-4z(1-z)m^2]\bigg],\label{f2lv}
\end{eqnarray}
which  leads in the zero momentum transfer $(q\rightarrow0)$ to 
\begin{eqnarray}
F_2(0)&=&\frac{\alpha}{2\pi}-\frac{11\alpha}{3\pi}\Bigg\{\frac{p.c^S.p}{m^2} \Bigg\},\label{f2lv0}
\end{eqnarray}
where $c^{S}$ is the symmetric part of the LV parameter $c_{\mu\nu}$ which is expected for a physical quantity.   Regarding (\ref{f2lv0}) some comments are in order: (i) the result obtained in  (\ref{f2lv0}) is valid only for $\frac{p.c^S.p}{m^2}\ll 1$.  For instance, the energy limits for $p_e\sim 100 TeV$,  $p_{\mu}\sim 200 TeV$, and $p_{\tau}\sim 50 GeV$ for the current bounds on $c_e\sim 10^{-16}$,  $c_{\mu}\sim 10^{-11}$, and $c_{\tau}\sim 10^{-8}$, respectively.   (ii) The LV correction which is obtained in  (\ref{f2lv0}) as a leading order correction is valid up to any order of $\alpha$.  It means that the leading order LV correction, which depends on the momentum of fermion  at the n-loop order, is 
\begin{eqnarray}
F_2(0)|_{n-loop}&=&F_2^{SM}|_{n-loop}-\frac{11\alpha}{3\pi}\Bigg\{\frac{p.c^S.p}{m^2} \Bigg\}.\label{f2lv0n}
\end{eqnarray}
Therefore, the anomalous magnetic moment of a charged fermion at the leading order of the LV parameter $c$ has been changed as 
\begin{eqnarray}
\delta a_{f}=F_2(0)|_{n-loop}-F_2^{SM}|_{n-loop}=\frac{11\alpha}{3\pi}\frac{p.c^S.p}{m^2},\label{amulv0n}
\end{eqnarray}
where the momentum dependence of $\delta a_{f}$ in (\ref{amulv0n}) would be interesting in high energy processes through the magnetic moment interaction.  Nevertheless, in the  experiments for measuring the anomalous magnetic moment of fermions where $p^2/m^2\ll 1$, the correction given in (\ref{amulv0n}) can be ignored.  Meanwhile, in the storage ring where the ‎muon is rotating ‎in the XY plane with $p^2/m^2\sim 10^3\gg 1$, the LV correction seems to be valuable.  To this end, we can consider $p_z=0$ and $\bar{p}_x=\bar{p}_y=0$ to find
\begin{eqnarray}
\delta a_{\mu}=8.5\times10^{-3}\{\frac{p_{0}^{2}c_{00}}{m_{\mu}^{2}}+\frac{p_{0}^{2}(c_{xx}+c_{yy})}{2m_{\mu}^{2}}\},
\label{a_NP_1}
\end{eqnarray}
where $p_0 \gg m_{\mu}$ is the muon energy in the ring.
In the standard sun-centered inertial frame \cite{suncenter}, the time and location dependence of the quantity $2c_{00}+c_{xx}+c_{yy}$ in the rotating frame can be obtained as follows:
‎\begin{eqnarray}‎
2c_{00}+c_{xx}+c_{yy}&=&2c_{TT}+(1-\sin^2\chi\cos^2\Omega‎
‎t)c_{XX}+(1-\sin^2\chi\sin^2\Omega‎
‎t)c_{YY}\nonumber\\
&+&\sin^2\chi c_{ZZ}-\frac{1}{2}\sin^2\chi\sin2\Omega t(c_{XY}+c_{YX})\nonumber\\‎
‎&-&\frac{1}{2}\sin2\chi\cos\Omega‎
‎t(c_{XZ}+c_{ZX})\nonumber\\‎
&-&\frac{1}{2}\sin2\chi\sin\Omega‎
‎t(c_{YZ}+c_{ZY}),
\label{a_NP_2}‎
‎\end{eqnarray}‎
‎where $\chi$ depends on the laboratory location.  ‎The time dependence in (\ref{a_NP_2}) leads to a day-night asymmetry in the ‎muon anomalous magnetic moment.  However, if such time-dependent experimental data are not readily available, one can average (\ref{a_NP_2}) on time which casts (\ref{a_NP_1}) into
\begin{eqnarray}
\delta a_{\mu}&=&8.5\times10^{-3}\dfrac{p_{0}^{2}}{2m_{\mu}^{2}}[(2c_{TT}+c_{XX}‎+c_{YY})‎ -‎\frac{1}{2}\sin^2\chi (c_{XX}‎+
c_{YY}-2 c_{ZZ})].\nonumber\\
\label{a_NP_5}
\end{eqnarray}
For example, in the E821 experiment the Brookhaven National Laboratory location is in $ \chi=49.1 $, $p_{0}\sim 3 GeV$ and $ m_{\mu}\sim 0.1 GeV$ \cite{E-821} lead to 
\begin{eqnarray}
\delta a_{\mu}=7.65\{c_{TT}+0.35(c_{XX}+c_{YY})+0.28c_{ZZ}\},
\label{a_NP_6}
\end{eqnarray}
which can explain the difference between the SM prediction and the experimental value for the muon anomalous magnetic moment if
\begin{eqnarray}
[c_{TT}+0.35(c_{XX}+c_{YY})+0.28c_{ZZ}]\sim 3.4\times 10^{-10}.
\label{a_NP_diff}
\end{eqnarray}
Meanwhile, for the available precision on the E821 experiment,
\begin{eqnarray}
[c_{TT}+0.35(c_{XX}+c_{YY})+0.28c_{ZZ}]<8.5\times 10^{-11},
\label{a_NP_7}
\end{eqnarray}
cannot affect the experimental value of $\mu$-AMDM.
\begin{table}[h]
	\begin{center}
		\caption{ LV bounds from $\mu$-AMDM in the muon storage ring experiments with energy $p_{0}=3 GeV$ for E821 and E989 and $p_{0}=0.32 GeV$ for J-PARC.}
		\label{tab:AMDM1}
		\scalebox{0.7}{
		\begin{tabular}{cccccc}
			\hline\hline
			Experiment & Precision(ppb) & $ \chi $ & LV components & Bound & Deviation\\
			\hline
			E821& 560 & 49.1 & $ c_{TT}+0.35(c_{XX}+c_{YY})+0.28c_{ZZ}$&$8.5 \times 10^{-11}$&$3.4\times 10^{-10}$ \\
			E989 & 140 & 48.2 & $c_{TT}+0.36(c_{XX}+c_{YY})+0.27c_{ZZ}$&$ 2.1\times 10^{-11}$ & $3.4\times 10^{-10}$\\
			J-PARC & 120 & 53.5 &  $c_{TT}+0.33(c_{XX}+c_{YY})+0.32c_{ZZ}$&$1.8\times 10^{-9}$ &$3.4\times 10^{-8}$\\
			\hline\hline
		\end{tabular}}
	\end{center}
\end{table}
For other current experiments such as E989 \cite{E989} and J-PARC(E34) \cite{JPARC}, bounds on the appropriate combination of the LV parameters are given in Table \ref{tab:AMDM1}.

\section{Conclusion}
We have considered QED part of SME to study the fermion-photon vertex up to the one loop level at the leading order of the LV parameter  $c_{\mu\nu}$ which preserve the CP symmetry.  Although $c_{\mu\nu}$ violates the particle Lorentz symmetry, it is a Lorentz tensor under the observer Lorentz transformation which leads to new form factors in the electromagnetic current; see (\ref{formfactor}) and (\ref{formfactor1}).  Meanwhile, all form factors depend on a new scalar as $p'.c.p$ which leads to a momentum dependence for the form factors even at the zero momentum transfer; see (\ref{f2lv0}).   In fact, at the high energy limit where $p^2\gg m^2$, the LV corrections on the form factors can be enhanced.   Such a correction has some new contribution on  the anomalous magnetic moment of a charged fermion as is given in  (\ref{a_NP_1}).  However,  the earth rotation, in addition to the location dependence, leads to a day-night asymmetry in the anomalous magnetic moment as is obtained in (\ref{a_NP_2}).    Furthermore,  we have obtained the time average of the obtained correction as is given in  (\ref{a_NP_5}).   Consequently, we have calculated the LV correction on the muon  anomalous magnetic moment in the storage ring.  
With the muon's energy, about $3$ GeV in the E821 experiment,  $[c_{TT}+0.35(c_{XX}+c_{YY})+0.28c_{ZZ}]=3.4 \times 10^{-10} $ can explain the current deviation between the experimental measurement and the theoretical prediction.  Nevertheless, in order not to have any observable effect on the  $\mu$-AMDM for the E821 experiment with $560$ ppb in precision, one finds a bound as  $8.5 \times 10^{-11}$ on $[c_{TT}+0.35(c_{XX}+c_{YY})+0.28c_{ZZ}]$; see (\ref{a_NP_7}).  Since the obtained correction in (\ref{a_NP_2}) depends on the location of laboratory where the measurement has been done, one can find different bounds on different combinations of the $c_{\mu\nu}$ components as is shown in Table \ref{tab:AMDM1}.  As the table shows for the future experiments the higher precision measurements lead to tighter bounds of about $2\times10^{-11}$. These  are the first bounds on the $c_{\mu\nu}$ components from the terrestrial experiment which are comparable with the astrophysical systems \cite{table}.

\section{Appendix A: Modified Gordon identities in QEDE}
Here, we introduce some useful identities in the QEDE which are modified by the $c$ parameter with respect to QED.  To this end, we begin with the Dirac equation in the SME as
\begin{eqnarray}
(\not \! p-m+c_{\mu\nu} p^{\nu}\gamma^{\mu})u(p)=0,
\end{eqnarray}
or
\begin{eqnarray}
\overline{u}(p')(\not \! p'-m+c_{\mu\nu} p'^{\nu}\gamma^{\mu})=0,
\end{eqnarray}
which can be cast into
\begin{eqnarray}
p^2u(p)=[m^2-2(p.c^S.p)]u(p),
\end{eqnarray}
and
\begin{eqnarray}
\overline{u}(p')p'^2=\overline{u}(p')[m^2-2(p'.c^S.p')],
\end{eqnarray}
or
\begin{eqnarray}
\overline{u}(p')q^2u(p)=\overline{u}(p')[2m^2-2(p'.c^S.p')-2(p.c^S.p)-2p.p']u(p).
\end{eqnarray}
Meanwhile, we can introduce the modified Gordon identity as follows:
\begin{eqnarray}
\bar{u}(p')(\gamma_{\mu}+c_{\nu\mu}\gamma^{\nu})u(p)=\bar{u}(p')\bigg (\frac{(p+p')_{\mu}}{2m}+i\frac{\sigma_{\mu\nu}q^{\nu}}{2m}\nonumber\\
+i\frac{c^{\alpha\nu}\sigma_{\mu\alpha}q_{\nu}+ic_{\alpha\mu}\sigma^{\alpha\nu}q_{\nu}}{2m}+\frac{(c_{\mu\nu}+c_{\nu\mu})(p+p')^{\nu}}{2m}\bigg)u(p).
\end{eqnarray}
\section{Appendix B: Magnetic form factor in the LV background $ c_{\mu\nu}$ at the one loop level}
Here, we give the detailed calculation of the form factor $F_2$ up to one loop level at the leading order of $c_{\mu\nu}$. The loop correction in the QEDE can be divided into two parts as  
\begin{eqnarray}
\delta\Gamma^{\mu}=\delta\Gamma^{\mu}_{SM}+\delta\Gamma^{\mu}_{LV},\label{gamma}
\end{eqnarray}
where $\delta\Gamma^{\mu}_{SM}$ shows the ordinary correction while $\delta\Gamma^{\mu}_{LV}$ is reserved for the LV part of the QEDE.  To find each part, first we focus on the $\mathcal{N}_7+\mathcal{N}_8$ where the Feynman parametrization is not as the usual one. By considering $k'=k+q$, the relations introduced in Appendix A and the LV parameter $c_{\mu\nu}$ as a traceless tensor, one finds
\begin{eqnarray}
\delta\Gamma^{\mu}_{\mathcal{N}_7+\mathcal{N}_8 }&=&\int\frac{d^{4}l}{(2\pi)^4}\int\frac{dxdydz\delta(x+y+z-1)}{(l^{2}-\Delta)^{4}}\nonumber\\
&&(12ie^{2})\mathcal{A}(c)\bar{u}(p')\bigg\{\gamma^{\mu}.[l^2-2(1-x)(1-y)q^2-2(1-4z+z^2)m^2]\nonumber\\
&&+i\frac{\sigma^{\mu\nu}q_{\nu}}{2m}.[-4z(1-z)m^2]\bigg\}u(p),\label{gamma78}
\end{eqnarray}
where $\mathcal{A}(c)$ is a function of the $c_{\mu\nu}$ tensor as follows
\begin{eqnarray}
\mathcal{A}(c)&=&[xy^2+y(1-y)^2].(q.c^S.q)+[yz^2].(q.c.p+p.c.q)\nonumber\\
&+&[(1-z)z^2].(p.c^S.p).
\end{eqnarray}
In contrast with $ \delta\Gamma^{\mu}_{\mathcal{N}_7+\mathcal{N}_8 } $, which contains only the LV contribution, $ \delta\Gamma^{\mu}_{\mathcal{N}_1+...+\mathcal{N}_6} $  has some contribution from both the ordinary SM and QEDE as 
\begin{eqnarray}
\delta\Gamma^{\mu}_{\mathcal{N}_1+...+\mathcal{N}_6}&=&\int\frac{d^{4}l}{(2\pi)^4}\int\frac{dxdydz\delta(x+y+z-1)}{(l^{2}-\Delta)^{3}}\nonumber\\
&&(-2ie^{2})\bar{u}(p')\bigg\{\gamma^{\mu}.[l^2-2(1-x)(1-y)q^2-2(1-4z+z^2)m^2]\nonumber\\
&&+i\frac{\sigma^{\mu\nu}q_{\nu}}{2m}.[-4z(1-z)m^2]\nonumber\\
&&\gamma^{\mu}.[-4z(1-y)(p'.c^S.p')-4z(1-x)(p.c^S.p)+8z(p'.c^S.p)\nonumber\\
&&+z(1+z)(P.c^S.P)+((x-y)^2-1-z)(q.c^S.q)]\nonumber\\
&&+c^{S\mu\nu}\gamma_{\nu}.[-4l^2+4xyq^2-4z(z-1)m^2]\nonumber\\
&&+c^{\nu\mu}\gamma_{\nu}.[l^2-2(1-x)(1-y)q^2-2(1-4z+z^2)m^2]\nonumber\\
&&-\frac{ic^{S\mu\nu}\sigma_{\nu\alpha}q^{\alpha}+ic^{S}_{\alpha\nu}\sigma^{\mu\nu}q^{\alpha}}{2m}.[8z(1-z)m^2]\nonumber\\
&&-\frac{ic_{\nu\alpha}\sigma^{\mu\nu}q^{\alpha}+ic^{\nu\mu}\sigma_{\nu\alpha}q^{\alpha}}{2m}.[4z(1-z)m^2]\nonumber\\
&&+c^{S}_{\alpha\beta}\gamma^{\alpha}.[P^{\mu}P^{\beta}(-2z(1-z))\nonumber\\
&&+2q^{\mu}q^{\beta}((x-y)^{2}+z-1)]\bigg\}u(p).\label{gamma16}
\end{eqnarray}
Therefore, comparing (\ref{gamma78}) and (\ref{gamma16}) with (\ref{gamma}) leads at the lowest order of $c_{\mu\nu}$ to
\begin{eqnarray}
\delta\Gamma^{\mu}_{SM}&=&\int\frac{d^{4}l}{(2\pi)^4}\int\frac{dxdydz\delta(x+y+z-1)}{(l^{2}-\Delta)^{3}}\nonumber\\
&&(-2ie^{2})\bar{u}(p')\bigg\{\gamma^{\mu}.[l^2-2(1-x)(1-y)q^2-2(1-4z+z^2)m^2]\nonumber\\
&&+i\frac{\sigma^{\mu\nu}q_{\nu}}{2m}.[-4z(1-z)m^2]\bigg\}u(p),
\label{SM}
\end{eqnarray}
and
\begin{eqnarray}
\delta\Gamma^{\mu}_{LV}&=&\int\frac{d^{4}l}{(2\pi)^4}\int\frac{dxdydz\delta(x+y+z-1)}{(l^{2}-\Delta)^{4}}\nonumber\\
&&(12ie^{2})\mathcal{A}(c)\bar{u}(p')\bigg\{\gamma^{\mu}.[l^2-2(1-x)(1-y)q^2-2(1-4z+z^2)m^2]\nonumber\\
&&+i\frac{\sigma^{\mu\nu}q_{\nu}}{2m}.[-4z(1-z)m^2]\bigg\}u(p)\nonumber\\
&&+\int\frac{d^{4}l}{(2\pi)^4}\int\frac{dxdydz\delta(x+y+z-1)}{(l^{2}-\Delta)^{3}}\nonumber\\
&&(-2ie^{2})\bar{u}(p')\bigg\{\gamma^{\mu}.[-4z(1-y)(p'.c^S.p')-4z(1-x)(p.c^S.p)\nonumber\\
&&+8z(p'.c^S.p)+z(1+z)(P.c^S.P)+((x-y)^2-1-z)(q.c^S.q)]\nonumber\\
&&+c^{S\mu\nu}\gamma_{\nu}.[-4l^2+4xyq^2-4z(z-1)m^2]\nonumber\\
&&+c^{\nu\mu}\gamma_{\nu}.[l^2-2(1-x)(1-y)q^2-2(1-4z+z^2)m^2]\nonumber\\
&&-\frac{ic^{S\mu\nu}\sigma_{\nu\alpha}q^{\alpha}+ic^{S}_{\alpha\nu}\sigma^{\mu\nu}q^{\alpha}}{2m}.[8z(1-z)m^2]\nonumber\\
&&-\frac{ic_{\nu\alpha}\sigma^{\mu\nu}q^{\alpha}+ic^{\nu\mu}\sigma_{\nu\alpha}q^{\alpha}}{2m}.[4z(1-z)m^2]\bigg\}u(p),
\label{LV}
\end{eqnarray}
where $P=p+p'$. It should be noted that (\ref{SM}) gives the exact contribution from the ordinary QED  at the one loop level.  Meanwhile, (\ref{LV}) indicates all contributions at the lowest order of $c_{\mu\nu}$ to the ordinary form factors and the new ones as well.  For instance, the magnetic form factor, which gives the anomalous magnetic dipole moment, is related to the coefficient of $i\frac{\sigma_{\mu\nu}q^{\nu}}{2m}$.  In fact, for the magnetic form factor, the finite part of the integrals in (\ref{SM}) and (\ref{LV}), at the zero momentum transfer and in the MMS scheme, leads to  
\begin{eqnarray}
F_2(0)&=&\frac{\alpha}{2\pi}-\frac{11\alpha}{3\pi}\Bigg\{\frac{p.c^S.p}{m^2} \Bigg\}.
\end{eqnarray}
\section{Appendix C: Electromagnetic form factors in the LV background $ c_{\mu\nu} $}
In this appendix, we examine all the form factors which are introduced in (\ref{formfactor}) up to the one loop level.  To this end, we compare (\ref{SM}) and (\ref{LV}) with (\ref{formfactor}).  For $F_1(q^2)$, which is the coefficient of $\gamma_{\mu}$, one has
\begin{eqnarray}
F_1(q^2)=F_1^{SM}(q^2)+F_1^{LV}(q^2),
\end{eqnarray}
where
\begin{eqnarray}
F_1^{SM}(q^2)&=&1+\int\frac{d^{4}l}{(2\pi)^4}\int\frac{dxdydz\delta(x+y+z-1)}{(l^{2}-\Delta)^{3}}\nonumber\\
&&(-2ie^{2})\bigg[l^2-2q^2(1-x)(1-y)-2(1-4z+z^2)m^2)\bigg],
\end{eqnarray}
is the ordinary electric form factor in the QED and 
\begin{eqnarray}
 F_1^{LV}(q^2)&=&\int\frac{d^{4}l}{(2\pi)^4}\int\frac{dxdydz\delta(x+y+z-1)}{(l^{2}-\Delta)^{3}}\nonumber\\
&&(-2ie^{2})\bigg[-4z(1-y)(p'.c^S.p')-4z(1-x)(p.c^S.p)+8z(p'.c^S.p)\nonumber\\
&&+z(1+z)(P.c^S.P)+((x-y)^2-1-z)(q.c^S.q)\bigg]\nonumber\\
&&+\int\frac{d^{4}l}{(2\pi)^4}\int\frac{dxdydz\delta(x+y+z-1)}{(l^{2}-\Delta)^{4}}\nonumber\\
&&(12ie^2)\bigg[\mathcal{A}(c)\times[l^2-2(1-x)(1-y)q^2-2(1-4z+z^2)m^2]\bigg],\nonumber\\
\end{eqnarray}
is the correction from the LV part of the Lagrangian.  However, by using an appropriate Wick rotation and performing the momentum integrals, one can easily show that
\begin{eqnarray} F_1^{LV}(q^2)&=&-\dfrac{\alpha}{4\pi}\int\frac{dxdydz\delta(x+y+z-1)}{\Delta}\nonumber\\
&&\bigg[-4z(1-y)(p'.c^S.p')-4z(1-x)(p.c^S.p)+8z(p'.c^S.p)\nonumber\\
&&+z(1+z)(P.c^S.P)+((x-y)^2-1-z)(q.c^S.q)\bigg]\nonumber\\
&&+\dfrac{\alpha}{\pi}\int\frac{dxdydz\delta(x+y+z-1)}{\Delta}\bigg[\mathcal{A}(c)\bigg]\nonumber\\
&&+\dfrac{\alpha}{2\pi}\int\frac{dxdydz\delta(x+y+z-1)}{\Delta^{2}}\nonumber\\
&&\mathcal{A}(c)\bigg[(-2(1-x)(1-y)q^2-2(1-4z+z^2)m^2)\bigg],\nonumber\\
\end{eqnarray}
and 
\begin{eqnarray}
 F_1^{LV}(0)=-\frac{2\alpha}{\pi}\frac{(p.c^S.p)}{m^2}[-1+\int dz\frac{1}{1-z}],
\end{eqnarray}
have not any UV divergences.  Therefore, the UV divergence for the electric form factor in the QEDE can be fixed similar to its counterpart in QED as follows:
\begin{eqnarray}
\delta F_1(q^{2})\rightarrow \delta F_1(q^{2})-\delta F_1(0),\label{F1}
\end{eqnarray}
where $\delta F_1=F_1-1$.  In fact, the electric charge normalization in this way can be fixed at zero momentum transfer.  Meanwhile, the IR divergence that appeared in the both SM and LV parts can be canceled by considering the soft bremsstrahlung amplitude ($M^{SB}_{LV}$) in the presence of the LV background as follows: 
\begin{eqnarray}
iM^{SB}_{LV}&=&\overline{u}(p')M_{SM}u(p)e[-\dfrac{p.\varepsilon^{*}}{p.k}
+\dfrac{p'.\varepsilon^{*}}{p'.k}]\nonumber\\
&&+\overline{u}(p')M_{SM}u(p)e[\dfrac{(p.c.p)(p.\varepsilon^{*})}{(p.k)^{2}}
+\dfrac{(p'.c.p')(p'.\varepsilon^{*})}{(p'.k)^{2}}]\nonumber\\
&&+\overline{u}(p')M_{SM}u(p)e[-\dfrac{(p.c.\varepsilon^{*})}{p.k}
+\dfrac{(p'.c.\varepsilon^{*})}{p'.k}]\nonumber\\
&&+\overline{u}(p')M_{SM}u(p)e[-\dfrac{c_{\alpha\beta}p^{\beta}\gamma^{\alpha}\gamma^{\mu}\varepsilon^{*}_{\mu}}{p.k}
+\dfrac{c^{\alpha\beta}p'^{\beta}\gamma^{\mu}\gamma^{\alpha}\varepsilon^{*}_{\mu}}{p'.k}],
\label{MSB}
\end{eqnarray}
where the first term in (\ref{MSB}) cancels the IR divergence of $F^{SM}_{1}(0)$ while the second term can fix the IR divergence of $F^{LV}_{1}(0)$.  It should be noted that, in contrast with the ordinary QED, the magnetic form factor and also the other new form factors have IR divergences as well.  In fact, the appearance of additional IR terms in (\ref{MSB}) are necessary for canceling the IR divergences in the other form factors.  For the magnetic form factor $ F_2$, (\ref{SM}) and (\ref{LV}) lead to
\begin{eqnarray}
F_2(q^2)&=& F_2^{SM}(q^2)+\int\frac{d^{4}l}{(2\pi)^4}\int\frac{dxdydz\delta(x+y+z-1)}{(l^{2}-\Delta)^{4}}\nonumber\\
&&(12ie^2)\bigg[\mathcal{A}(c)[-4z(1-z)m^2]\bigg],
\end{eqnarray}
where
\begin{eqnarray}
F_2^{SM}(q^2)&=&\int\frac{d^{4}l}{(2\pi)^4}\int\frac{dxdydz\delta(x+y+z-1)}{(l^{2}-\Delta)^{3}}(-2ie^{2})\nonumber\\
&& \times \bigg[4z(z-1)m^2\bigg].
\end{eqnarray}
Therefore, after  performing the momentum integrals and at the zero momentum transfer one has
\begin{eqnarray}
F_2(0)&=&\frac{\alpha}{2\pi}+\frac{2\alpha}{\pi}\frac{(p.c^S.p)}{m^2}\int dz\frac{z^3}{1-z}\nonumber\\
&=&\frac{\alpha}{2\pi}-\frac{11\alpha}{3\pi}\Bigg\{\frac{p.c^S.p}{m^2}\Bigg\}+\frac{2\alpha}{\pi}\frac{(p.c^S.p)}{m^2}\int dz\frac{1}{1-z},\label{F2}
\end{eqnarray}
where the last term shows the IR divergence of $F_2(0)$. As is already mentioned,  (\ref{MSB}) has additional IR terms which can resolve the IR part of the other form factors in the LV case. For this purpose the last term of (\ref{MSB}) can be rewritten as
\begin{eqnarray}
\overline{u}(p')M_{SM}u(p)e[\dfrac{(\varepsilon^{*}.c.q)}{2k}
-\dfrac{ic_{\alpha\beta}\varepsilon^{*}_{\mu}\sigma^{\alpha\mu}(p+p')^{\beta}}{8k}],
\end{eqnarray}
 which can remove the IR divergence of $ F_2 $ at the one loop level.
\\
Finally, $ F_c $ can be derived from (\ref{SM}) and (\ref{LV}) as follows:
\begin{eqnarray}
F_c(q^2)&=& \int\frac{d^{4}l}{(2\pi)^4}\int\frac{dxdydz\delta(x+y+z-1)}{(l^{2}-\Delta)^{3}}\nonumber\\
&&(-2ie^2)\bigg[-4l^2+4xyq^2+16z(1-z)m^2\bigg].
\end{eqnarray}
Nevertheless, we do not have any physical interpretation for the form factors that appear in the $ F_c $ which can be cast into 
\begin{eqnarray}
F_c(0)=-\frac{2\alpha}{\pi}+IR,\label{Fc}
\end{eqnarray}
at the zero momentum transfer.

\end{document}